\documentclass{aa}  

\usepackage{graphicx}
\usepackage{txfonts}
\usepackage[breaklinks,pdfnewwindow=true,colorlinks=true,citecolor=blue]{hyperref}
\usepackage[version=3]{mhchem}
\usepackage{xspace}
\usepackage{orcidlink}
\newcommand{\orcid}[1]{\unskip\protect\href{https://orcid.org/#1}{\protect\includegraphics[width=8pt,clip]{logo_orcid}}}

\newcommand{\jybm}{\ensuremath{\rm Jy\,beam^{-1}}\xspace}
\newcommand{\msun}{\ensuremath{\rm M_\odot}\xspace}

\newcommand{\cc}{\ensuremath{\rm cm^{-3}}\xspace}

\newcommand{\kms}{{\ensuremath{{\rm km\, s^{-1}}}}\xspace}

\usepackage{xcolor}

\begin{document}

   \title{Probing the physics of star formation (ProPStar)
   \thanks{Based on observations carried out under 
   project number S21AD with the IRAM NOEMA Interferometer and 
   090-21 with the IRAM 30-m telescope. 
   IRAM is supported by INSU/CNRS (France), MPG (Germany) and IGN (Spain)}}

  \subtitle{I. First resolved maps of the electron fraction and cosmic-ray ionization rate {in NGC 1333}}

   \author{
        Jaime E. Pineda
        \inst{1}\orcidlink{0000-0002-3972-1978}
        \and
        Olli Sipil\"a\inst{1}\orcidlink{0000-0002-9148-1625}
        \and
        Dominique M. Segura-Cox\inst{2,1,}\thanks{NSF Astronomy and Astrophysics Postdoctoral Fellow}\orcidlink{0000-0003-3172-6763}
        \and
        Maria Teresa Valdivia-Mena\inst{1}\orcidlink{0000-0002-0347-3837}
        \and
        Roberto Neri\inst{3}\orcidlink{0000-0002-7176-4046}
        \and
        Michael Kuffmeier\inst{4,1}\orcidlink{0000-0002-6338-3577}
        \and
        Alexei V. Ivlev\inst{1}\orcidlink{0000-0002-1590-1018}
        \and
        Stella S. R. Offner\inst{2}\orcidlink{0000-0003-1252-9916}
        \and
        Maria Jose Maureira\inst{1}\orcidlink{0000-0002-7026-8163}
        \and
        Paola Caselli\inst{1}\orcidlink{0000-0003-1481-7911}
        \and
        Silvia Spezzano\inst{1}\orcidlink{0000-0002-6787-5245}
        \and
        Nichol Cunningham\inst{5}\orcidlink{0000-0003-3152-8564}
        \and
        Anika Schmiedeke\inst{1}\orcidlink{0000-0002-1730-8832}
        \and
        Mike Chen\inst{6}\orcidlink{0000-0003-4242-973X}
        }

\institute{
    Max-Planck-Institut f\"ur extraterrestrische Physik, Giessenbachstrasse 1, D-85748 Garching, Germany\\
    \email{jpineda@mpe.mpg.de}
    \and
    Department of Astronomy, The University of Texas at Austin, 2500 Speedway, Austin, TX 78712, USA
    \and
    Institut de Radioastronomie Millim\'etrique (IRAM), 300 rue de la Piscine, F-38406, Saint-Martin d'H\`eres, France
    \and
    Department of Astronomy, University of Virginia, Charlottesville, VA, 22904, USA
    \and
    IPAG, Universit\'{e} Grenoble Alpes, CNRS, F-38000 Grenoble, France
    \and
    Department for Physics, Engineering Physics and Astrophysics, Queen's University, Kingston, ON, K7L 3N6, Canada
    }

   \date{Received Month Day, Year; accepted Month Day, Year}

 
  \abstract
   {Electron fraction and cosmic-ray ionization rates in star-forming regions are 
   important quantities in astrochemical modeling
   and are critical to the degree of coupling between neutrals, ions, and electrons, which regulates the dynamics of the magnetic field.
   However, these are difficult quantities to estimate.}
   {We aim to derive the electron fraction and cosmic-ray
   ionization rate maps of an active star-forming region.}
   {%
   We combined observations of the nearby NGC 1333 star-forming region carried out with the NOEMA interferometer 
   and IRAM 30-m single dish to generate high spatial dynamic range maps of different molecular transitions. 
   We used the \ce{DCO+} and \ce{H^{13}CO+} ratio (in addition to complementary data) to  
   estimate the electron fraction and produce cosmic-ray ionization rate maps.}
   {We derived the first large-area electron fraction and cosmic-ray ionization rate resolved maps in a star-forming region, with typical values of $10^{-6.5}$ and $10^{-16.5}$ s$^{-1}$, 
   respectively.
   The maps present clear evidence of enhanced values around embedded young stellar objects (YSOs). 
   This provides strong evidence for locally accelerated cosmic rays.
   We also found a strong enhancement toward the northwest region in the map that 
   might be related either to an interaction with a bubble or to locally generated 
   cosmic rays by YSOs.
   We used the typical electron fraction and derived a 
   magnetohydrodynamic (MHD) turbulence dissipation 
   scale of 0.054 pc, which could be tested with future observations.
   }
   {We found a higher cosmic-ray ionization rate compared to the canonical value 
   {for $N(\ce{H2})=10^{21}-10^{23}$ cm$^{-2}$}
   of $10^{-17}$ s$^{-1}$ in the region, 
   and it is likely generated by the accreting YSOs. 
   The high value of the electron fraction suggests that new disks will 
   form from gas in the ideal-MHD limit.
   This indicates that local enhancements 
   of $\zeta(\ce{H2})$, due to YSOs, should be taken into account in the analysis 
   of clustered star formation.}


   \maketitle
%

\section{Introduction}
Stars form in cold, dense cores within molecular clouds \citep{Pineda2023-PP7}.  
Probing these regions enables observations of the initial conditions for star and disk formation. 
Some of the key aspects still poorly constrained in such processes are the electron fraction, $X(e)$,  
and cosmic-ray ionization rate, $\zeta (\ce{H2})$.
Both of these quantities play an important role in astrochemical models \citep{Ceccarelli2023-PP7} and 
in the coupling between magnetic fields and dense gas 
\citep{Pineda2021-B5_Ions_Neutral,Pattle2023-PP7_Bfield,Tsukamoto2023-PP7_Disk_Formation}.

In the case of a constant cosmic-ray ionization rate, the electron fraction should follow 
a relation with a density of 
$X(e)\propto n^{-0.5}$ \citep{McKee1989-Electron_Fraction,Caselli_1998-Ion_Fraction}. 
The normalization of the relation depends substantially on whether the gas shows no depletion 
of metals \citep{McKee1989-Electron_Fraction}, appropriate for a low-density environment, 
or if the depletion of metals is taken into account in the modeling 
\citep{Bergin_Langer_1997-Core_Chemical_Model,Caselli_1998-Ion_Fraction}, appropriate for denser regions. 

Observationally, it is not possible to directly measure the electron fraction. Instead, it must be inferred from the combined analysis of several molecules. 
Toward dense cores, different works have derived typical values in the dense regions of 
$X(\ce{e}) \approx 10^{-8}-10^{-6}$
\citep{Guelin1977-DCOp_Ionization,Caselli_1998-Ion_Fraction,Bergin1999-Ions_Massive_Cores,Maret2007-B68_Ions}, 
while in 
the transition from the diffuse to the dense medium in TMC1 
($n(\ce{H2})\sim 10^3$ \cc), an electron fraction of  $X(e) \approx 9.8 \times 10^{-8} - 3.6 \times 10^{-7}$ has been found
\citep{Fuente2019-GEMS_I,Rodriguez-Baras2021-GEMS_IV}.
Recently, high angular resolution ALMA observations of the protostellar core B335 have provided an 
estimate of the electron fraction, $X(\ce{e}) \approx 1 - 8\times 10^{-6}$,
within 1000 au \citep{Cabedo2023-B335_Ions}.
This last work showed a radial variation of the electron fraction, with increased values toward the protostar.

Similarly, the cosmic-ray ionization rate is estimated 
toward low column density line of sights 
by observing the \ce{H3+} absorption feature toward background sources, 
which provide a value of $\zeta(\ce{H2}) \approx 2\times 10^{-16}\,{\rm s}^{-1}$ in diffuse molecular gas \citep{Indriolo2012-CR_Diffuse_Clouds,Neufeld2017-CRIR,Padovani2020-CosmicRays_Star_Formation}. 
In comparison, the ionization rate calculated using the particle spectra 
measured by the Voyager spacecrafts is 10 times lower  
$\zeta(\ce{H2}) \approx 10^{-17}\,{\rm s}^{-1}$ \citep{Cummings2016-CosmicRays_Voyager1} 
{at column densities 
between $10^{21}$--$10^{23}$ cm$^{-2}$ \citep[see][]{Padovani2022-Cosmic_Rays_JWST}, 
suggesting some degree of variation throughout the Galaxy.
Toward higher column densities, other methods are required, and therefore, different molecular ratios and/or 
detailed modeling are used to derive these values. 
Theoretical models propose that the local value of the 
cosmic-ray ionization rate can be increased 
by the accretion process onto the protostar \citep{Padovani+2016,GachesOffner2018}
over the "standard"  cosmic-ray ionization rate often adopted for  
molecular clouds,  $\zeta(\ce{H2}) \approx 10^{-17}\,{\rm s}^{-1}$  \citep{Padovani+2009}.
Observations of the OMC-2 FIR4 region have revealed enhanced values of the cosmic-ray 
ionization rate \citep{Ceccarelli2014-OMC2-FIR4_CR,Fontani2017,Favre2018}.
Similarly, ALMA observations of the protostellar core B335 have also suggested an enhanced value of the 
cosmic-ray ionization rate \citep{Cabedo2023-B335_Ions}.
These works show that a better understanding of the role of protostars and their surroundings is important to 
properly understand the star- and disk-formation processes.

Isolated cores, such as L1544 and B68, provide a great opportunity to study in detail the physical conditions of star and disk formation
\citep{Maret2007-B68_Ions,Redaelli2021-L1544_Ions}. 
However, more clustered environments provide a chance to study the mode of star formation where most stars form, 
although at the price of a more difficult environment to model.
The NGC 1333 region in the Perseus cloud, at a distance of $\approx300$ pc \citep{zucker_2018}, 
offers a good opportunity to study an active star-forming region in a clustered environment
\citep{Kirk2006-SCUBA_Perseus,Winston2010-Chandra_NGC1333_Serpens,Gutermuth2008-NGC1333_Spitzer}.
Notably, studies on the dense gas in the region, as traced by \ce{NH3} or \ce{N2H+}, have revealed narrow 
filamentary structures with a coherent velocity  \citep{Friesen2017-GAS_DR1,Hacar2017-NGC1333,Dhabal2019-NGC1333_VLA,M_Chen+20}.

In this work, we present the first results of the Probing the physics of star formation (ProPStar) survey, which attempts to study 
the physical conditions of star-forming regions with interferometric observations in order to
connect the molecular cloud and disk scales.
We present new observations of the NGC 1333 region, which allowed us to 
present the first resolved large-area maps of the 
electron fraction and cosmic-ray ionization rate of a low-mass star-forming region. 
We compare the electron fraction and cosmic-ray ionization rate maps 
with theoretical expectations and explore possible explanations for the observed 
spatial variations. 
Finally, we discuss the implications of the results.

\section{Data}
In the following, we summarize the single dish and interferometric observations 
for the \ce{H^{13}CO+} and \ce{DCO+} lines. 
The whole NGC 1333 star-forming region is shown in Fig.~\ref{fig:NGC1333}, 
and the coverage area of the combined observations is marked 
by a dashed box, which includes the SVS 13 and NGC1333 IRAS 4 systems.

\begin{figure}[ht]
    \centering
    \includegraphics[width=\columnwidth]{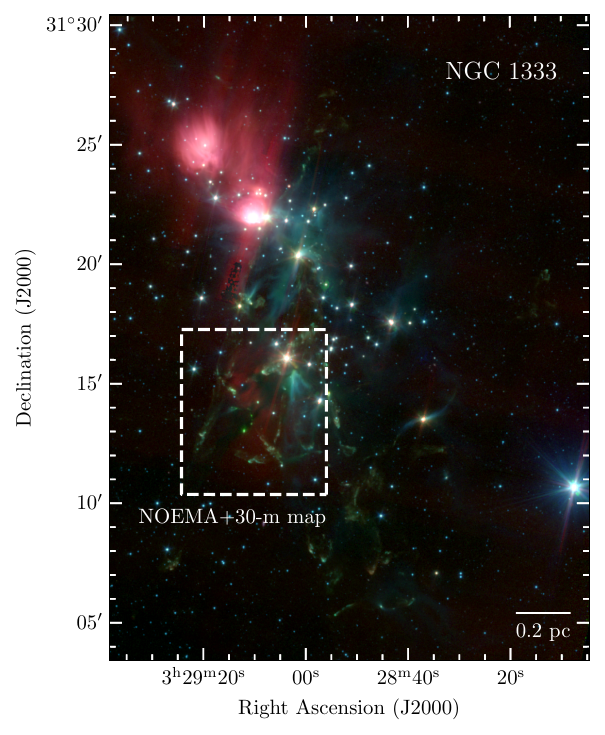}
    \caption{Color image of the NGC 1333 region using {\it Spitzer}
    IRAC 3.6, 4.5, and 8.0 $\mu$m.
    The image highlights the presence of outflows in a 
    green color.
    The area covered by the combined NOEMA and 30-m observations 
    is shown with the dashed box.
    The scale bar is shown in the bottom-right corner.}
    \label{fig:NGC1333}
\end{figure}

\subsection{IRAM 30-m telescope}
The observations were carried out with the IRAM 30-m telescope at Pico Veleta (Spain) on 
2021 November 9, 10, and 11 and on
2022 February 19 and 20 under project 091-21.
The EMIR E090 receiver and the FTS50 backend were employed. 
We used two spectral setups to %
cover the \ce{H^{13}CO+} (1--0) and \ce{DCO+} (1--0)  
lines at 72.0 and 86.7 GHz (see Table \ref{tab:spectral_setup}).
We mapped a region of $\approx$150\arcsec$\times$150\arcsec with the On-the-Fly mapping technique and using position switching.
The data reduction was performed using the CLASS program of the GILDAS package.\footnote{\url{http://www.iram.fr/IRAMFR/GILDAS}} 
The beam efficiency, $B_{eff}$, was obtained using the Ruze formula 
(available in CLASS), and it was used to convert the observations 
into main beam temperatures, $T_{\rm mb}$.
The noise for the \ce{H^{13}CO+} (1--0) and \ce{DCO+} (1--0) 
cubes is 48 and 79 mK, in $T_{\rm mb}$ scale, respectively. 

\subsection{NOEMA interferometer}
The observations carried out with the IRAM NOrthern Extended Millimeter Array (NOEMA) interferometer within the S21AD program 
using the Band 1 receiver
were obtained on 2021 
July 18, 19, and 21; 
August 10, 14, 15, 19, 22, and 29; and 
September 1
in the D configuration.
We observed a total of 96 pointings, which were separated 
into four different scheduling blocks.
The mosaic's center is located at 
$\alpha_{J2000}$=03$^{\rm h}$29$^{\rm m}$10.2$^{\rm s}$, 
$\delta_{J2000}$=31
$\degr$13$\arcmin$49.4$\arcsec$.
We used the PolyFix correlator with a LO frequency of 82.505\,GHz and an instantaneous
bandwidth of 31 GHz spread over two sidebands (upper and lower) and two polarizations.
The centers of the two 7.744\,GHz-wide sidebands were separated by 15.488\,GHz. 
Each sideband is composed of two adjacent basebands of $\sim$3.9\,GHz width (inner and outer basebands). 
In total, there are thus eight basebands that were fed into the correlator. 
The spectral resolution is 2\,MHz throughout the 15.488\,GHz effective bandwidth per polarization. 
Additionally, a total of 112 high-resolution chunks were placed, 
each with a width of 64\,MHz and a fixed
spectral resolution of 62.5\,kHz. 
Both polarizations (H and P) are covered with the same spectral setup, and therefore the high-resolution chunks provide 66 dual polarization spectral windows.
The high spectral resolution windows used in this work are listed in Table~\ref{tab:spectral_setup}.

\subsection{Image combination}
We resampled the original 30-m data to match the spectral setup 
of the NOEMA observations. %
We used the task \verb+uvshort+ to generate the pseudo-visibilities 
from the 30-m data for each NOEMA pointing. 
The imaging was done with natural weighting, a support mask, and 
using the SDI deconvolution algorithm.

\begin{table}
\caption{Lines observed in the high-spectral resolution band.\label{tab:spectral_setup}}
\begin{tabular}{lccc}
\hline \hline
Transition & Rest Freq. & Beam\tablefootmark{a} & rms \\
 & (MHz) & (\arcsec$\times$\arcsec) & (m\jybm)\\
 \hline
%
\ce{DCO+} (1--0) & 72039.3122 & 6.2$\times$6.0(-37$\degr$) & 23\\
\ce{H^{13}CO+} (1--0) & 86754.2884 & 5.0$\times$4.9(-43$\degr$)  & 15\\
\hline
\end{tabular}
\tablefoot{
\tablefoottext{a}{Beam size before smoothing and matching to common beam and grid.}}
\end{table}

The \ce{H^{13}CO+} (1--0) was then convolved to match the 
\ce{DCO+} (1--0) beam size using the \verb+spectral-cube+ and 
\verb+RadioBeam+ Python packages. 
The noise level of the combined images is reported in Table~\ref{tab:spectral_setup}. 
Both cubes were converted to units of K, and the integrated intensity
maps were calculated between 5.4 and 10 \kms, which covers all the 
emission seen in both molecules. 
The integrated intensity maps are shown in Fig.~\ref{fig:maps_TdV}.

\begin{figure*}[ht]
    \centering
    \includegraphics[width=0.49\textwidth]{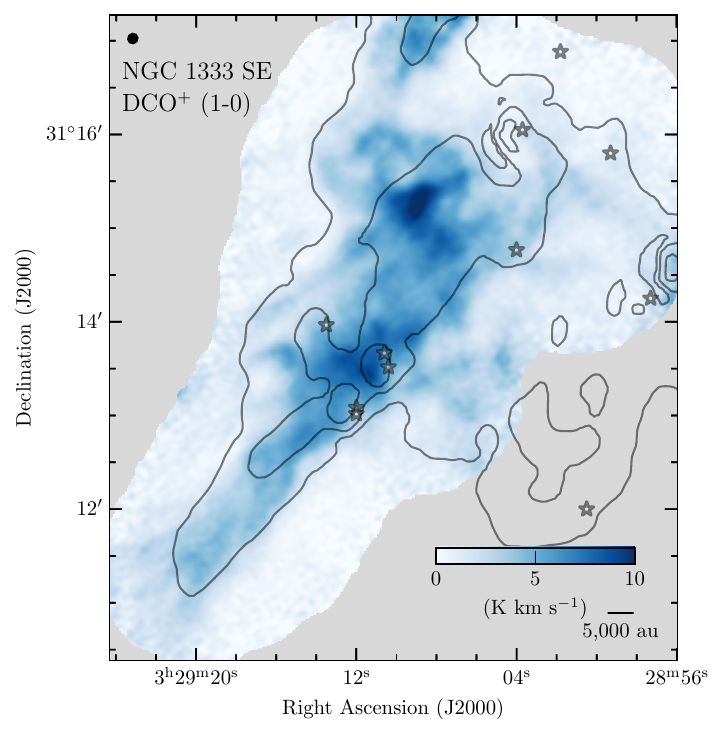}
    \includegraphics[width=0.49\textwidth]{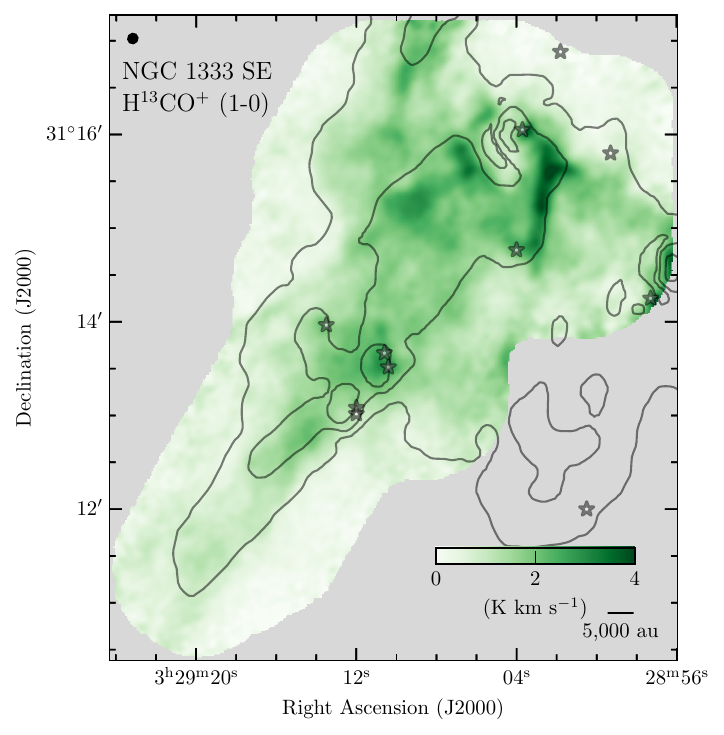}
    \caption{Maps of the \ce{DCO+} and \ce{H^{13}CO+} (1--0) emission obtained with NOEMA and 30-m telescopes.
    The beam size and scale bar are shown in the top left and bottom right, respectively.
    The contours correspond to the {\it Herschel}-based 
    \ce{H2} column map; the first level is at  $\log_{10}(N(\ce{H2}))=22.2$, and the step between levels 
    is 0.5 dex.
    The stars mark the positions of the YSOs identified by \cite{Dunham2015-YSO_c2d} using {\it Spitzer} observations.
    }
    \label{fig:maps_TdV}
\end{figure*}

\subsection{\ce{H2} column density}
We used the total \ce{H2} column density, $N(\ce{H2})$, 
derived using the 
{\it Herschel} observations \citep{Pezzuto2021-Perseus_Herschel}, which are available in the {\it Herschel} Gould Belt Survey repository.\footnote{\url{http://www.herschel.fr/cea/gouldbelt/en/Phocea/Vie_des_labos/Ast/ast_visu.php?id_ast=66}}
The effective angular resolution of the $N(\ce{H2})$
maps is 18.2\arcsec.
The map was regridded to match the \ce{DCO+} integrated intensity
map. 
The resulting map is shown in Fig.~\ref{fig:Herschel}.

\begin{figure}
    \centering
    \includegraphics[width=0.49\textwidth]{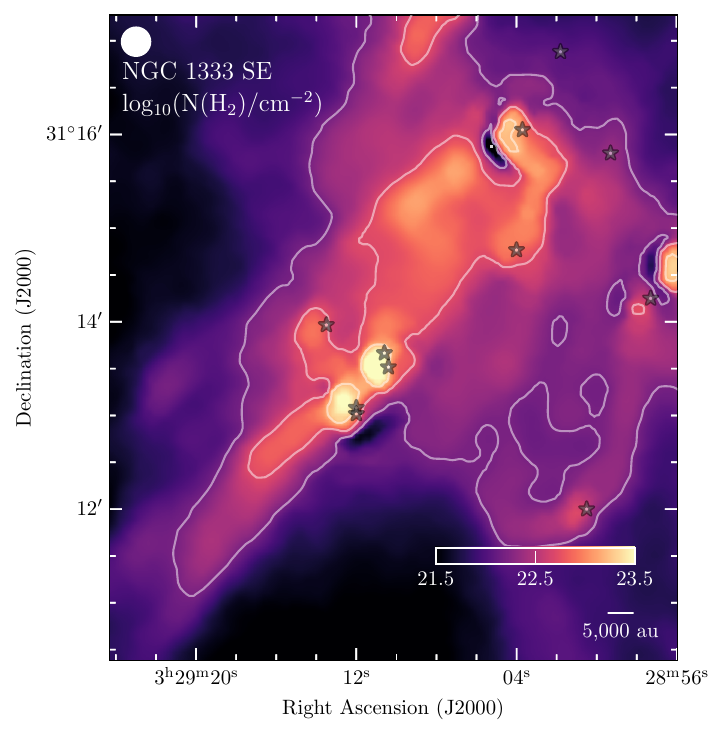}
    \caption{Map of the {\it Herschel}-based \ce{H2} column map, 
    $N(\ce{H2})$.
    The contours are drawn starting at $\log_{10}(N(\ce{H2}))=22.2$ and 
    with a step between levels of 0.5 dex. These contours 
    are the same as in Fig.~\ref{fig:maps_TdV}.
    The beam size and scale bar are shown in the top left and bottom right, respectively.}
    \label{fig:Herschel}
\end{figure}

\subsection{JCMT \ce{C^{18}O} observations}

We used the \ce{C^{18}O} (3--2) observations of NGC 1333 
taken with HARPS at JCMT \citep{Curtis2010-JCMT_Perseus_CO}.
The angular resolution of these observations is 17.7\arcsec, and the main beam efficiency is 0.66 
\citep[as used in][]{Curtis2010-JCMT_Perseus_CO}.
We smoothed the \ce{C^{18}O} data to match the 
angular resolution of the {\it Herschel}-based 
$N(\ce{H2})$. 
The integrated intensity map we calculated is between 6 and 10 \kms. 
This range covers all the  emission seen in the cube. 

\begin{figure}
    \centering
    \includegraphics[width=\columnwidth]{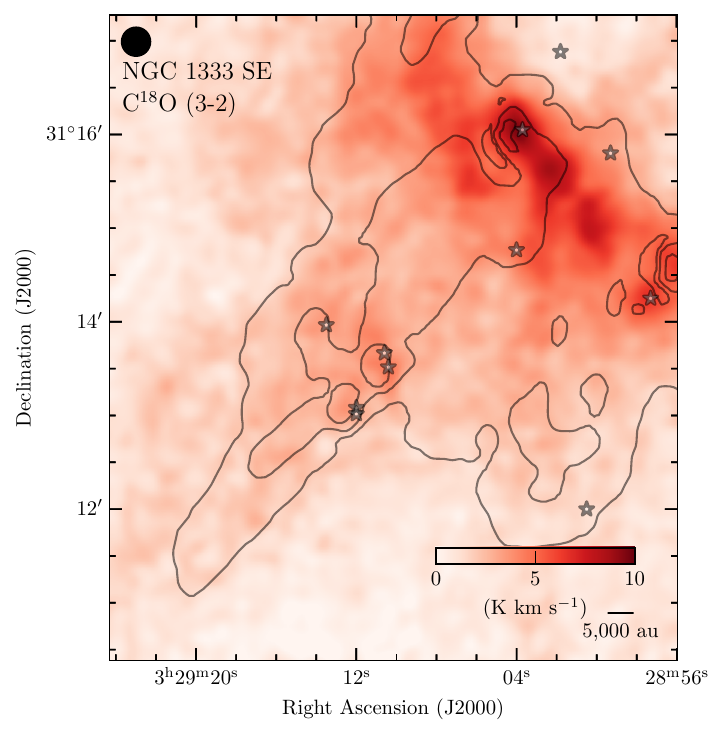}
    \caption{Integrated intensity map of the \ce{C^{18}O} (3--2)
    transition line.
    The contours correspond to the {\it Herschel}-based 
    \ce{H2} column map, with the same contours as in Fig.~\ref{fig:maps_TdV}.
    The beam size and scale bar are shown in top left and bottom right, respectively.}
    \label{fig:C18O}
\end{figure}

\section{Analysis}
\subsection{Estimate of ionization fraction and 
cosmic-ray ionization rate}

One of the most commonly used ionization fraction tracers is
[\ce{DCO+}]/[\ce{HCO+}] \citep{Guelin1977-DCOp_Ionization,Guelin1982-Ionization_Fraction,Dalgarno1984-D_Fractionation,Caselli_1998-Ion_Fraction}.
We estimated the ionization fraction, $X(e)$, and 
the cosmic-ray ionization rate, $\zeta(\ce{H2})$, following 
\cite{Caselli_1998-Ion_Fraction}. 
The method uses the main reaction paths for the formation 
and destruction for \ce{HCO+} and \ce{DCO+}. 
Although other line ratios and techniques have been explored \citep{Bron2021-Ions_Models},
we used a method that combines the bright lines 
available in the observations and the \ce{CO} depletion.
The analysis relations between the observables and 
the desired physical parameters are
\begin{eqnarray}
    X(e) &=& \frac{2.7\times 10^{-8}}{R_{\rm D}} - 
    \frac{1.2 \times 10^{-6}}{f_D}~, \label{eq:x_e}\\
    \zeta(\ce{H2}) &=& \left[7.5\times 10^{-4} X(e) + 
    \frac{4.6\times 10^{-10}}{f_D}\right] 
    X(e)\, n(\ce{H_2})\, R_{\rm H}~, \label{eq:zeta}
\end{eqnarray}
where 
$R_{\rm D} \equiv [\ce{DCO+}]/[\ce{HCO+}]$; 
$f_D \equiv [\ce{^{12}CO}]/[\ce{H2}]/ \left([\ce{^{12}CO}]/[\ce{H2}]\right)_{\rm fiducial}$; 
$\left([\ce{^{12}CO}]/[\ce{H2}]\right)_{\rm fiducial}$ is the expected \ce{^{12}CO} abundance; 
$R_{\rm H} \equiv [\ce{HCO+}]/[\ce{^{12}CO}]$;  
and $n(\ce{H2})$ is the average \ce{H2} number density.

However, since \ce{HCO+} is usually optically thick and it also 
traces outflow emission (not traced by \ce{DCO+}), 
we used \ce{H^{13}CO+} and the canonical 
isotropic ratio of $[\ce{^{12}C}]/[\ce{^{13}C}]=68$ \citep{Milam2005-12C_13C_Relation} 
to derive $R_{\rm D}$. 
Similarly, we used \ce{C^{18}O} and the canonical 
isotropic ratio of $[\ce{^{12}CO}]/[\ce{C^{18}O}]=560$  
\citep{Wilson1994-Review_Abundance_ISM} 
to estimate the \ce{^{12}CO} column density.

\subsection{Column densities}
\subsubsection{\ce{DCO+} and \ce{H^{13}CO+}}
We used the optically thin approximation to derive the column densities \citep{Mangum2015-Column_Density}. 
In this case, we used the following expression to calculate 
the total column densities:
\begin{equation}
    N_{\rm tot} =  
    \frac{8\pi\,\nu^3}{c^3} \frac{Q(T_{\rm ex})}{A_{\rm ul} 
    g_{\rm u}\,e^{-E_{\rm up} / h\,T_{\rm ex}}}
    \frac{1}{\left(e^{h \nu/k_{B}T_{\rm ex}} - 1\right)} 
    \frac{\int T_{\rm mb} dv}{J(T_{\rm ex}) 
    - J(T_{\rm bg})}~,
\end{equation}
where $\nu$ is the frequency of the transition observed, 
$A_{ul}$ is the Einstein coefficient for the transition 
from level $u$ to $l$,
$g_{\rm u}$ is the degeneracy of level $u$, 
$E_{\rm up}$ is the energy of level $u$, 
$T_{\rm ex}$ is the excitation temperature,
$Q(T)$ is the partition function at temperature $T$,
and 
$J(T)\equiv h\nu/k_B\, (\exp(h\nu/k_B\,T) - 1)$ is the 
Rayleigh-Jeans equivalent temperature. 
We used the $A_{\rm ul}$, $g_{\rm u}$, and $E_{\rm up}$ listed 
in the LAMDA database,
and implemented in the \verb+molecular_columns+ 
package.\footnote{\url{https://github.com/jpinedaf/molecular_columns}}

In this work, we used a constant excitation temperature of 10~K for 
both transitions. Moreover, we only used estimates of the column density 
toward pixels with a signal-to-noise ratio of the line 
above 7.5 ($T_{\rm peak} > 7.5 \times rms$) in order to obtain 
robust column densities.

\subsubsection{\ce{C^{18}O}}
The \ce{C^{18}O} column density was derived using the 
optically thin approximation, 
\begin{equation}
    N(\ce{C^{18}O}) = 5\times 10^{12}\, \frac{T_{\rm ex}}{\rm K} \, \exp{(31.6\, {\rm K}/T_{\rm ex})} \, \frac{\int T_{\rm mb} {\rm d}v}{\kms} ~{\rm cm^{-2}}~,
\end{equation}
and an excitation temperature of $T_{\rm ex}=$12 K, 
which were previously used by 
\cite{Curtis2010-JCMT_Perseus_CO}.
These excitation temperature values are similar to those 
derived using \ce{^{12}CO} (1--0) 
but with lower angular resolution in this region \citep{Pineda2008-Perseus_CO}.
We also estimated the optical depth in the map for the assumed $T_{\rm ex}$
and obtained a typical value of $\sim$0.4, which suggests that the optical 
depth is not an issue across the map. However, it might 
produce lower limits to the derived $N(\ce{C^{18}O})$. 
{Appendix \ref{appendix:tau} shows details regarding the optical depth calculation 
and a figure showing the map.}

\subsection{\ce{CO} depletion factor}
We estimated the \ce{CO} depletion factor, $f_D$, using the \ce{C^{18}O} 
column density, $N(\ce{C^{18}O})$, the isotopologue abundance ratio of 
$\ce{^{12}CO}/\ce{C^{18}O}=560$ \citep{Wilson1994-Review_Abundance_ISM}, 
the \ce{H2} column density derived from {\it Herschel}, 
and the expected $[\ce{^{12}CO}]/[\ce{H2}] = 2.7\times 10^{-4}$ 
value, as determined in Taurus \citep{Punanova2022-CO_CH3OH_Abundance} 
and comparable to the value of $2.5\times 10^{-4}$ obtained in NGC1333 
at a lower angular resolution \citep{Pineda2008-Perseus_CO}.
When comparing the $N(\ce{H2})$ and $N(\ce{C^{18}O})$ column 
densities maps, it was clear that the \ce{H2} map traces more 
extended emission, and therefore, an offset is present,
\begin{equation}
    N(\ce{H2}) = N(\ce{C^{18}O}) \frac{[\ce{H2}]}{[\ce{C^{18}O}]}
    + \delta_{18}~.
\end{equation}
This offset represents the \ce{H2} column in the molecular cloud 
that is not traced by the \ce{C^{18}O} (3--2) emission, and therefore 
it needed to be removed from the abundance calculations.
We estimated the offset, $\delta_{18}=2.7\times 10^{21}$ cm$^{-2}$,
as the median value of 
the difference observed assuming an undepleted abundance in the 
column density range of 
$3\times 10^{21}\,{\rm cm^{-2}} < N(\ce{H2}) < 5\times 10^{21}\,{\rm cm^{-2}}$.
This is equivalent to determining the offset 
required to achieve undepleted abundances at the lower column 
densities in the region.
The resulting map is shown in Fig.~\ref{fig:fd_map}, with 
depletion fractions increasing at higher column densities.

\begin{figure}[ht]
    \centering
    \includegraphics[width=0.49\textwidth]{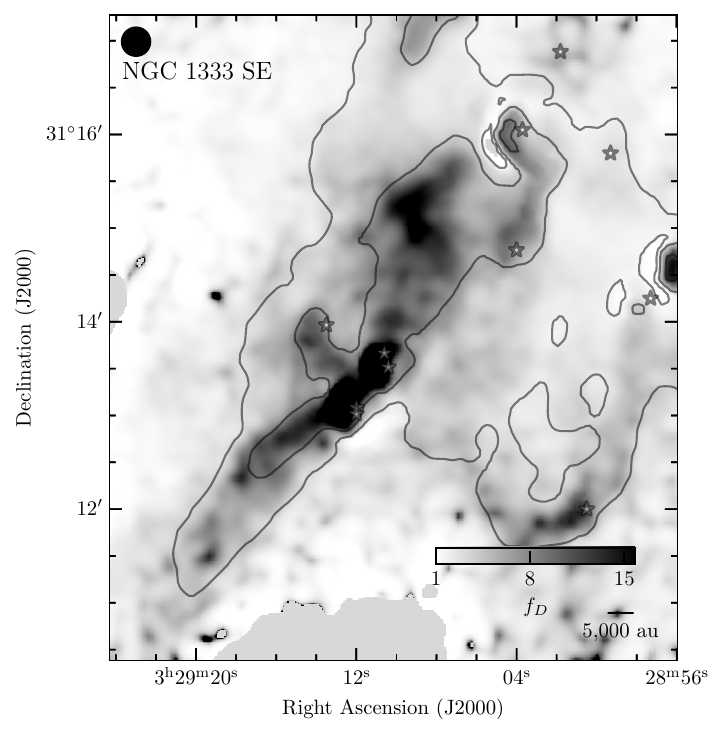}
    \caption{Map of the \ce{CO} depletion fraction, $f_D$.
    The beam size and scale bar are shown in the top-left and bottom-right corners, respectively.
    The contours correspond to the {\it Herschel}-based 
    \ce{H2} column map; the contours are the same as in Fig.~\ref{fig:maps_TdV}.
    }
    \label{fig:fd_map}
\end{figure}

\subsection{Volume density}
We estimated the volume density as
\begin{equation}
    n(\ce{H2}) = \frac{N(\ce{H2}) - \delta_{18}}{\delta L}~,
    \label{n_H}
\end{equation}
where the $N(\ce{H2})$ is the \ce{H2} column density, 
$\delta_{18}$ is the derived offset with \ce{C^{18}O} ,
and $\delta L$ is the depth of the structure. 
We assumed $\delta L=0.4$ pc ($\sim 60\times 10^3$ au), 
which is the depth for \ce{HCO+} derived for the nearby L1451 
\citep{Storm2016-CLASSY_L1451}. 
The corresponding volume density map is shown in Fig.~\ref{fig:n_H2_map}, 
and it displays densities similar to those constrained 
from \ce{^{12}CO} and \ce{^{13}CO} (1--0) observations \citep{Pineda2008-Perseus_CO}. 
The typical density in places with a detection of \ce{DCO+} and \ce{H^{13}CO+} is $\approx 4\times 10^3$ \cc,
which suggests that this density is likely a lower limit since this region is bright in
\ce{NH3} \citep{Friesen2017-GAS_DR1,Dhabal2019-NGC1333_VLA} and \ce{N2H+} \citep{Hacar2017-NGC1333}, 
which are reliable high-density tracers.

\begin{figure}[ht]
    \centering
    \includegraphics[width=0.49\textwidth]{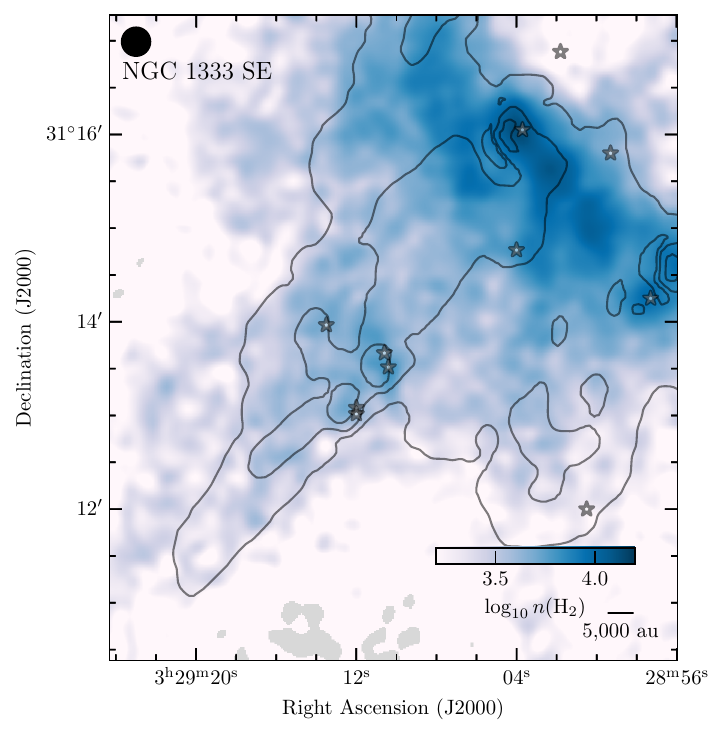}
    \caption{Volume density map estimated from {\it Herschel} 
    \ce{H2} column density.
    The beam size and scale bar are shown in the top left and bottom right, respectively.
    The contours correspond to the {\it Herschel}-based 
    \ce{H2} column map; the contours are the same as in Fig.~\ref{fig:maps_TdV}.
    }
    \label{fig:n_H2_map}
\end{figure}

\subsection{Electron fraction\label{analysis:xe}}
We used the relation of Eq.~(\ref{eq:x_e}) together with 
the previously presented measurements to estimate the 
electron fraction across the active star-forming region,
and it is presented in the left panel of Fig.~\ref{fig:Xe}.
The underlying distribution of the electron fraction 
in the region was estimated using the 
kernel density estimate (KDE), and it is shown in the 
right panel of Fig.~\ref{fig:Xe}.
A typical value (median of the $\log_{10}$) of $\langle X(e)\rangle = 10^{-6.5}$ 
was derived from these measurements.
{We estimate that the error associated with using the wrong excitation temperature 
(in the range of 5 to 30 K)
in the column density determination yields an uncertainty of less than 10\% in $R_D$ and $X(e)$.}

\begin{figure*}[ht]
    \centering
    \includegraphics[width=0.49\textwidth]{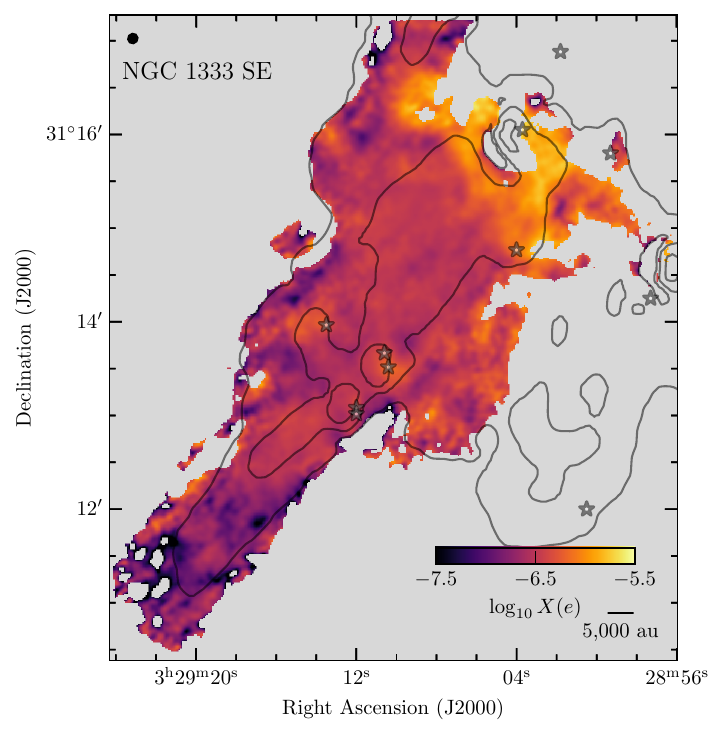}
    \includegraphics[width=0.49\textwidth]{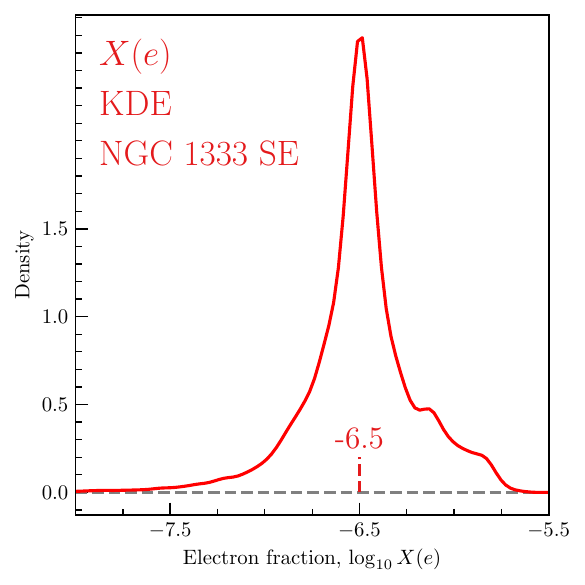}
    \caption{Derived electron fraction in the region.
    Left: Map of the electron fraction, $X(e)$. 
    The map shows a smooth distribution but with a clear enhancement in the 
    northwest section.
    The beam size and scale bar are shown in the top left and bottom right, respectively.
    The contours correspond to the {\it Herschel}-based 
    \ce{H2} column map; the contours are the same as in Fig.~\ref{fig:maps_TdV}.
    Right: Kernel density estimate of the $X(e)$ 
    distribution across the mapped region. 
    A typical value (median of the $\log_{10}$) of $10^{-6.5}$ 
    was derived from these measurements and is 
    marked with a red dashed vertical line.
    }
    \label{fig:Xe}
\end{figure*}

\subsection{Cosmic-ray ionization rate}
We used the relation of Eq.~(\ref{eq:zeta}) together with 
the previously presented measurements to estimate the 
cosmic-ray ionization rate across the active star-forming region,
and the estimate is presented in the left panel of Fig.~\ref{fig:zeta}.
The underlying distribution of the cosmic-ray ionization rate
in the region was estimated using the 
KDE, and it is shown in the 
right panel of Fig.~\ref{fig:zeta}.
We derived a typical value (median of the $\log_{10}$) of 
$\langle\zeta(\ce{H2})\rangle=10^{-16.3}$ s$^{-1}$ 
from these measurements.
{We estimate that the error associated with using the wrong excitation temperature 
(with the temperature within the range of 5 and 30 K)
$R_H$ could be underestimated by up to a factor of two, 
in addition to the uncertainty of less than 10\% in $R_D$ and $X(e)$ mentioned in Sect.~\ref{analysis:xe}.
This indicates that the $\zeta(\ce{H2})$ could be underestimated by a factor of two toward hot regions ($T_{ex}>20$K),
but elsewhere, these uncertainties are of the order 10--30\%.}

\begin{figure*}[ht]
    \centering
    \includegraphics[width=0.49\textwidth]{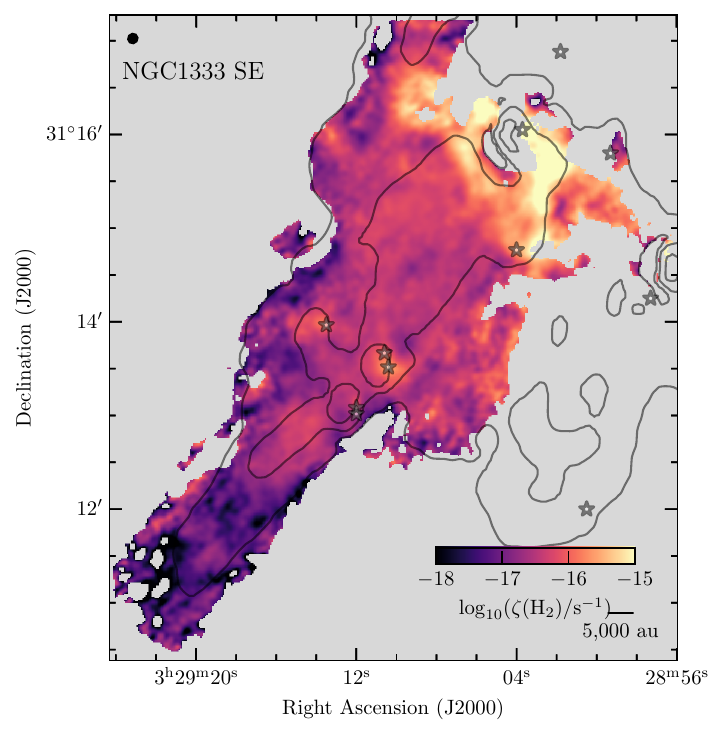}
    \includegraphics[width=0.49\textwidth]{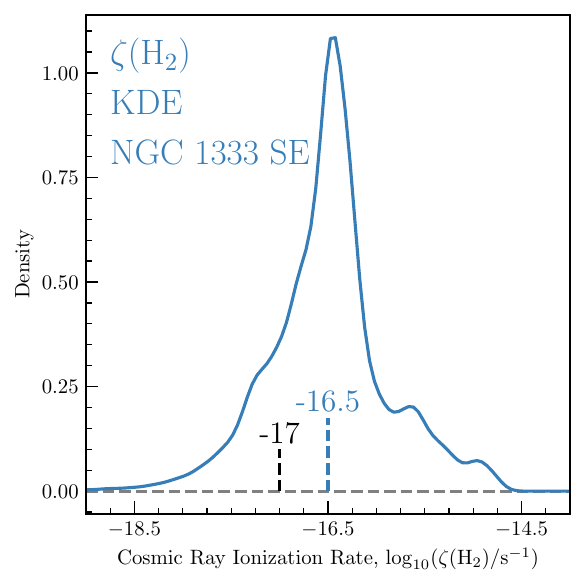}
    \caption{Derived cosmic-ray ionization rate in the region.
    Left: Map of the cosmic-ray ionization rate, $\zeta(\ce{H2})$. 
    The map shows a smooth distribution but with the value clearly increasing 
    in the northwest section.
    The beam size and scale bar are shown in top left and bottom right, respectively.
    The contours correspond to the {\it Herschel}-based 
    \ce{H2} column map; the contours are the same as in Fig.~\ref{fig:maps_TdV}.
    Right: Kernel density estimate of the $\zeta(\ce{H2})$ 
    distribution across the mapped region.
    The median value of the $\log_{10} \zeta(\ce{H2}) / {\rm s}^{-1}=-16.5$
    and the canonical value of $10^{-17}$ s$^{-1}$ are 
    marked with the blue and black dashed vertical lines, respectively.
    }
    \label{fig:zeta}
\end{figure*}

\section{Discussion}
\subsection{Electron fraction and cosmic-ray ionization rate distributions}

The spatial distribution for both of the main parameters estimated
in this work, $X(e)$ and $\zeta(\ce{H2})$, are shown in the left panel of 
Figs. \ref{fig:Xe} and \ref{fig:zeta}.
These maps show the first resolved maps of these quantities, 
which present little substructure, except close to the 
cloud or structure's edge and close to the young protostars. 
{These maps were derived at the angular resolution of the \ce{DCO+} 
map, but we note that since $n(\ce{H2})$ and $f_D$ were derived using 
{\it Herschel} data, we assumed that they smoothly vary between 16$\arcsec$ and $\approx$6$\arcsec$, 
which might be less accurate toward the young protostars. }

\subsection{Electron fraction as a function of density}

Several attempts have been carried out to 
parametrize the variation of the 
electron fraction as a function of density. 
\cite{McKee1989-Electron_Fraction} presented an analytic relation for 
the electron fraction as a function of density assuming  
\emph{no depletion of metals and a constant cosmic-ray ionization rate}. 
{This relation is expected to work even at high densities \citep{McKeeOstriker07}.}
After the depletion of metals was identified as an 
important process in the denser regions within star-forming regions, 
an updated electron fraction relation with the density relation 
was presented by \cite{Caselli2002-L1544_Ions}, where  
a constant cosmic-ray ionization rate was assumed. 
Further detailed modeling of the prestellar core
L1544 using different molecular transitions
constrained the radial profile electron fraction \citep{Redaelli2021-L1544_Ions}.
This derived relation is closer to that 
derived by \cite{Caselli_1998-Ion_Fraction}, 
while \citep{McKee1989-Electron_Fraction} provides a clear overestimation.

We compared the KDE distribution of the values determined from the data
and the analytic relations in Fig.~\ref{fig:Xe_n_H2}. In addition,
we added a comparison with the relation obtained from modeling the 
L1544 prestellar core \citep{Redaelli2021-L1544_Ions}. 
A decrease in the electron fraction, $X(e)$, at higher densities, $n(\ce{H2})$, 
is a common trend from the analytical (and modeling) results.
Notably, Fig.~\ref{fig:Xe_n_H2} shows a consistent increase 
in the electron fraction toward higher density areas, while 
beyond $n(\ce{H2})\gtrsim 10^{3.7}$ \cc, the relation steepens.
This steeper relation is mostly driven by the northwest 
area in the map, which is coincident with a previously suggested 
bubble interacting with the cloud 
\cite[e.g.,][]{Dhabal2019-NGC1333_VLA,DeSimone2022-SiO_Fingers}.

\begin{figure}[ht]
    \centering
    \includegraphics[width=\columnwidth]{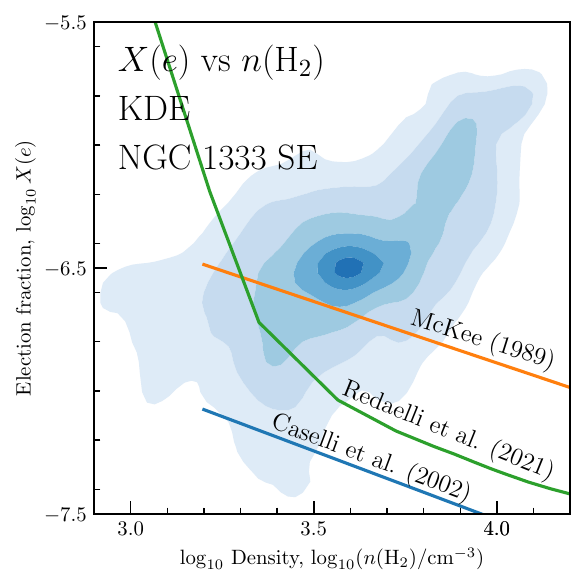}
    \caption{Electron fraction, $X(e)$, as a function of \ce{H2} density. 
    The KDE of the underlying distribution was obtained from the data.
    The contour levels are drawn starting at 0.5$\sigma$ and progressing outward in steps of 0.5$\sigma$, 
    where the $\sigma$ levels are equivalent to that of a bivariate normal distribution. %
    The analytic relations from \cite{McKee1989-Electron_Fraction} and 
    \cite{Caselli2002-L1544_Ions} as well as the 
    result from chemical modeling of L1544 \citep{Redaelli2021-L1544_Ions} 
    are shown with orange, blue, and green curves, 
    respectively.
    The analytic relations shown here do not take into account 
    the local generation of cosmic rays.
    }
    \label{fig:Xe_n_H2}
\end{figure}

A key difference between the analytical relations 
for electron fraction as a function of density and data 
is the assumption of no local sources of cosmic rays. 
Recently, different works have suggested that enhanced levels of 
cosmic rays 
\citep[thanks to their local generation by YSOs, e.g.,][]{Ceccarelli2014-OMC2-FIR4_CR,Padovani+2016} 
could provide a higher value of electrons than previously expected. 
This important difference might be the main physical reason as to why 
the analytical relations have such a poor match to the data, 
even in the case of \cite{McKee1989-Electron_Fraction}, who reported a similar 
predicted value at the typical density ($\approx 10^{3.7}$ \cc) even though 
there is a clear measurement of the depletion of metals ($f_D$).

The derived densities are comparable to previous 
estimates.
The mean density in the region
probed with \ce{^{12}CO} and \ce{^{13}CO} (1--0) transitions 
are $\approx 10^3$ \cc \citep{Pineda2008-Perseus_CO}.
Therefore, although our density estimate is rather simple 
and likely a lower limit, 
it should not be the main reason for the disagreement between the 
measured electron fractions and the theoretical models.

\subsection{Local generation of cosmic rays}

Recently, several indirect observations have suggested an extremely high ionization rate over $10^{-15}$~s$^{-1}$ in protostellar environments
\citep[e.g.,][]{Ceccarelli2014-OMC2-FIR4_CR, Fontani2017, Favre2018,Cabedo2023-B335_Ions}. 
Also, synchrotron emission, the fingerprint of the presence of relativistic electrons, has been detected in the shocks that develop along protostellar jets \citep[e.g.,][]{Beltran2016, Rodriguez-Kamenetzky2017, Osorio2017, Sanna2019}. 
Neither signature can be explained by interstellar cosmic rays 
since they are strongly attenuated at the high column densities 
typical of protostellar environments. 
As argued by \citet{Padovani+2016} and \citet{Padovani2021-CR_Model_Application}, shocks forming along the jets and on the protostellar surface could be the regions where the locally accelerated cosmic rays are produced.

The left panel of Fig.~\ref{fig:zeta} shows a dramatic local increase of $\zeta(\ce{H2})$ occurring in the northwest (upper) section of the map. Furthermore, there is a noticeable enhancement of $\zeta(\ce{H2})$ when in proximity to the embedded YSOs, which are also in the central area of the studied region. 
In fact, the volume density in dense cores around YSOs 
(typically within $\lesssim$0.1~pc, \citealt{Kirk2006-SCUBA_Perseus}) 
is much higher than the "large-scale" estimate based on Eq.~(\ref{fig:n_H2_map}). 
Therefore, according to Eq.~(\ref{eq:zeta}), the actual value of $\zeta(\ce{H2})$ is 
expected to be further drastically enhanced around these YSOs. 
In order to gain insights into the underlying mechanisms of the observed phenomenon, 
in Fig.~\ref{fig:zeta_Bfield_outflow}, we have overlaid the cosmic-ray ionization rate map 
with the outflow orientation of the different embedded sources (see Table~\ref{tab:Rdisk}), 
the dust polarization vectors obtained from the 
JCMT SCUBA2-POL Bistro Survey \citep{Ward-Thompson2017-BISTRO} on this 
region \citep{Doi2020-NGC1333_BISTRO_Bfield, Doi2021-NGC1333_BISTRO_Filaments_Pinched}, and 
the X-ray sources detected with Chandra \citep{Winston2010-Chandra_NGC1333_Serpens}.

\begin{figure}[ht]
    \centering
    \includegraphics[width=\columnwidth]{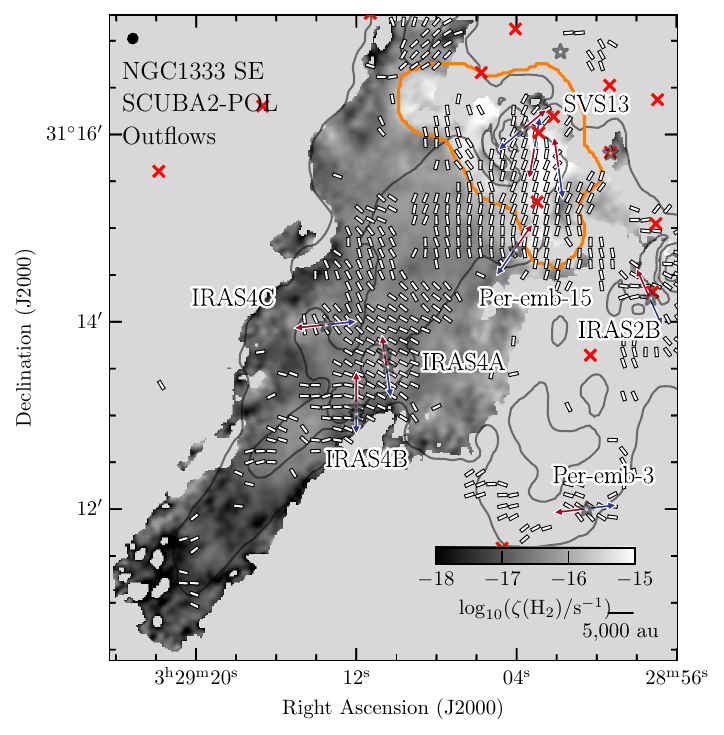}
    \caption{Outflows and magnetic field orientation over $\zeta(\ce{H2})$.
    The background image is the derived $\zeta(\ce{H2})$ map, 
    as shown in Fig.~\ref{fig:zeta}. 
    The region identified with a local cosmic-ray ionization rate 
    enhancement is marked with an orange contour.
    The outflow lobes (as reported in previous works) are 
    shown by red and blue arrows. 
    The magnetic field orientation, measured from continuum polarization observations with SCUBA2-POL, 
    are shown by the segments of white lines.
    The X-ray sources are marked with red crosses.
    The beam size and scale bar are shown in the top-let and bottom-right corners, respectively.}
    \label{fig:zeta_Bfield_outflow}
\end{figure}

First, we ruled out the X-ray sources as a possible explanation of the observed ionization enhancement. 
A significant attenuation of X-ray photons produced by 
source emitting in the keV energy range starts at column densities of $\sim10^{23}$~cm$^{-2}$ \citep{Igea1999}, which is about the measured maximum value of $N(\ce{H2})$ 
in the upper section (where most of the sources are located). 
Therefore, {X-rays would result in a relatively isotropically enhanced} 
$\zeta(\ce{H2})$ around the sources, which is clearly incompatible 
with the distribution of $\zeta(\ce{H2})$ seen in this region, 
revealing no correlation with the source locations.

However, we have strong indicators that the sources of the 
ionization enhancement are the young embedded objects. 
The first obvious fact is the noticeable increase of $\zeta(\ce{H2})$ around most of the YSOs. 
Given that we strongly underestimated the gas density in the proximity to these objects, 
the actual values of $\zeta(\ce{H2})$ are expected to be much higher than those shown in Fig.~\ref{fig:zeta}. 
Second, the elongated region with the highest local ionization rate of 
$\gtrsim10^{-15}$~s$^{-1}$ in the upper section of the map is approximately 
parallel to the magnetic field lines, practically connecting the SVS13 system and Per-emb-15. 
If these objects are able to generate energetic charged particles, 
their further propagation and the resulting ionization must occur along the local field lines \citep{Fitz-Axen2021-Cosmic_Rays_Turbulent_Cores}.  
We note that in this work, we also include a figure of the difference between the observed CRIR and the 
expected values \citep{Padovani2018-Cosmic_Ray_Models} in Appendix~\ref{Appendix:delta_zeta}, which shows the same trends.

{We compared the CRIR as a function of the typical YSO distance, calculated 
as the harmonic mean of the individual distances,\footnote{This is akin as to deriving the typical flux 
at a given distance}
\[
\frac{1}{d_{YSO}^2} = \frac{1}{N} \sum_{i=1}^N \frac{1}{d_i^2}~,
\]
where $d_{i}$ is the distance to the different YSOs marked in Fig.~\ref{fig:zeta_Bfield_outflow}.
Figure~\ref{fig:CRIR_dYSO} shows the KDE for the map, and it shows that within 15,000 au,
the CRIR is relatively constant, while there is a reduction in the CRIR at larger distances.
This is evidence that the YSOs are related to the increase of the CRIR. 
Although, we cannot confirm a correlation between YSO luminosity and CRIR 
due to the small sample size and the issues determining accurate CRIR toward the YSOs.}

\begin{figure}[h]
\includegraphics[width=\columnwidth]{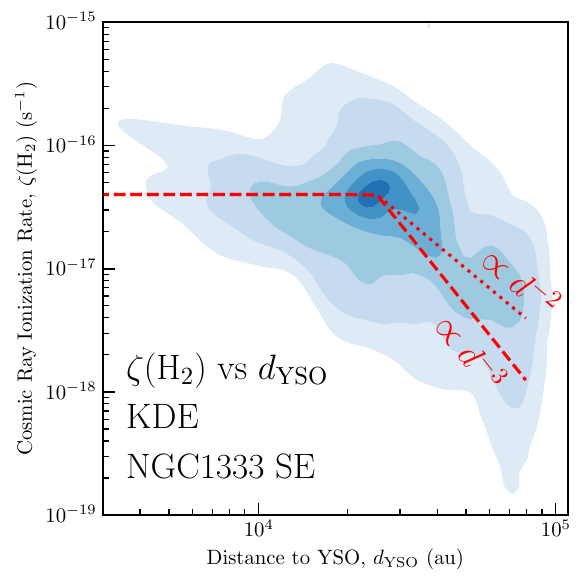}
\caption{{Cosmic-ray ionization rate as a function of distance to YSOs.
The KDE of the CRIR and the typical distance to the YSOs, $d_{YSO}$, was calculated over the map.
The dotted and dashed lines show the inner constant section, and the two different power-law 
decays help guide the eye.
\label{fig:CRIR_dYSO}}
}
\end{figure}

It is worth noting that the three-color {\it Spitzer} image in Fig.~\ref{fig:NGC1333} 
shows a strong outflow emission (green) from the 
SVS13 region, which overlaps with a fraction of the high enhancement 
in the cosmic-ray ionization rate. 
The SVS13 region hosts several strong and well-studied outflows 
\citep{Maret2009-NGC1333_Outflows_H2,Hodapp2014-SVS13_Outflows,Lefevre2017-CALYPSO_SVS13A_Jets,Dionatos2020-NGC1333_Outflows_SVS13}.
It is possible that part of the local enhancement in 
CRIR is related to the outflow cavity, especially if we 
are probing the emission closer to the outflow cavity itself.
{Models following CR propagation indicate that the CR distribution near accelerating protostellar sources is likely inhomogeneous with the CR enhancement occurring along the direction of the outflow \citep{Fitz-Axen2021-Cosmic_Rays_Turbulent_Cores}.}

We would like to emphasize that most of the observed region (in particular, the lower section of the map in Fig.~\ref{fig:zeta}) is dominated by $\zeta(\ce{H2}) \sim10^{-17}$~s$^{-1}$. This level of ionization, likely generated by interstellar cosmic rays, is comparable to the "standard" value of the ionization rate in diffuse molecular clouds \citet{Glassgold1974-zeta}. 
This value is also in good agreement with constraints on the CRIR derived from energy arguments. For the Milky Way, \citet{Krumholz2023-Cosmic_Rays_Galaxy_Budget} inferred a mean primary ionization budget due to star-formation activity of $\zeta(\ce{H2}) \sim 2-5 \times 10^{-17}$~s$^{-1}$, significantly below the value measured in diffuse molecular gas. 
These values can be reconciled if CR acceleration sources are distributed inhomogeneously.

Remarkably, we can independently estimate the expected upper bound of the ionization rate near NGC 1333 produced by interstellar cosmic rays in the surrounding diffuse gas. One of the measurements of \ce{H3+} ions reported by \citet{Indriolo2012-CR_Diffuse_Clouds} was performed in the direction of the target star HD 21483. Along this line of sight, the measurements probe the cosmic-ray ionization at the same radial distance of 300~pc, which is only 4--5 pc away from NGC 1333 in the plane of sky. The resulting non-detection yields a $1\sigma$ upper limit of $9\times10^{-17}$~s$^{-1}$ for the ionization rate. Given that the obtained spectra of \ce{H3+} lines show no visible traces of absorption (see their Fig.~10), we conclude that the actual value of $\zeta(\ce{H2})$ in this region is substantially lower and may well be close to the expected "standard" value. 

In a separate publication, we will present a detailed analysis 
of the ionization rate map. We will also perform a quantitative comparison 
of the results with available theories of propagation of interstellar 
and locally accelerated cosmic rays.

\subsection{Evidence of shocks and interaction}
The spatial distribution for both of the main parameters estimated in this work, 
$X(e)$ and $\zeta(\ce{H2})$, are shown in the left panel of Figs. \ref{fig:Xe} and \ref{fig:zeta}.
Both quantities present a clear one-order-of-magnitude enhancement 
toward the northwest in the mapped region. 
This could be consistent with the suggestion that the region is affected by 
interaction with an expanding shell \citep{DeSimone2022-SiO_Fingers} coming 
from the northwest section of the map (the SVS13 region).
A similar interaction, although at a much larger scale, is suggested in the nearby 
Taurus cloud with the Local Bubble and the Per-Tau shell \citep{Soler2023-Taurus_MC_Compression},
which could also be occurring in the Perseus Cloud.

A second region along the southwest of the map has previously been 
suggested as showing evidence of interaction between 
the cloud and a turbulent cell \citep{Dhabal2019-NGC1333_VLA} or 
of a second bubble impacting the cloud \citep{DeSimone2022-SiO_Fingers}.
However, we do not see clear evidence in $X(e)$ or $\zeta(\ce{H2})$ 
of any of these features, although we cannot rule them out.

\subsection{The relevance of nonideal-MHD effects}

On scales of a couple of beams away from the embedded sources,
the ionization fraction is $\approx 10^{-6.5}$ 
(see Fig.~\ref{fig:Xe}). 
It is usually assumed that at a relatively high  
electron fraction, the nonideal-MHD terms are negligible.
Studying the initial core collapse phase \cite{Wurster2018-ionization_disks} 
showed that above cosmic-ray ionization rates of 
$\langle \zeta(\ce{H2}) \rangle > 10^{-14} {\rm s}^{-1}$, the
results are practically indistinguishable from results 
obtained by assuming zero resistivity (i.e., ideal MHD).
The values measured in NGC 1333 are slightly higher with 
respect to the often-assumed canonical value for 
prestellar cores of $\langle \zeta(\ce{H2}) \rangle \approx 10^{-17} {\rm s}^{-1}$
\citep{SpitzerTomasko1968, UmebayashiNakano1990}, 
and therefore,
the nonideal-MHD effects are expected to not play an important role in current and future disk formation. 

\subsection{MHD wave cutoff}

In a molecular cloud where the ionization level is low, the Alfv\'e waves cannot propagate when the collision frequency between ions and neutrals is comparable to or smaller than the MHD wave frequency. 
Therefore, a critical length for wave propagation beyond which waves do not transmit can be defined.
The critical length for wave propagation 
\citep{StahlerPalla-2005,Mouschovias2011-MHD} 
is obtained from
\begin{equation}
\lambda_{min} = 
\sqrt{\frac{\pi}{4 \mu \, m_{\ce{H}} \, n(\ce{H2})}} \frac{B_o}{n(\ce{H2})\, X(e)\, 
\langle \sigma\, v \rangle~}, 
\end{equation}
where $B_o$ is the unperturbed magnetic field and 
$\langle \sigma\, v \rangle$ is the rate coefficient for elastic collisions.
The rate coefficient term is approximated by the Langevin term,
\begin{equation}
\langle \sigma\, v \rangle = 1.69\times 10^{-9}\, {\rm cm^3\,s^{-1}}~,
\end{equation}
for \ce{HCO+}--\ce{H2}  collisions 
\citep{McDaniel:1973vs}. 
We estimated the magnetic field strength, $B_o$, using the 
relation of \cite{Crutcher2010-B_field_Bayesian}, 
a typical volume density of $10^{3.6}$ \cc, 
and the median value of $X(\ce{e})=10^{-6.5}$.
With these values, we then obtained an MHD wave cutoff scale 
of 0.054 pc (11\,100 au). 
Therefore, we predicted that if the turbulence is 
MHD in nature, then a dissipation scale should be observed at 
$\approx$0.054 pc (or $\approx$37$\arcsec$ at the distance of Perseus).

Previously, this scale was proposed as a 
possible origin for the transition to coherence 
between the supersonic cloud and subsonic cores 
\citep[$\sim$0.1 pc;][]{Goodman1998-Transition_to_Coherence,Caselli2002-Cores_N2Hp,Pineda2010-B5_Transition_to_Coherence}. 
Recent observations of dense gas with different 
tracers (e.g., \ce{N2H+} and \ce{NH3}) and 
angular resolutions showed that the southern 
end filament presents subsonic levels of 
turbulence at scales of $\approx$40$\arcsec$
\citep{Friesen2017-GAS_DR1,Hacar2017-NGC1333,Dhabal2019-NGC1333_VLA,Sokolov2020-NGC1333_Bayesian},
which corresponds to 0.054 pc, or 11\,200 au, at the distance 
of Perseus.
This is consistent with the estimated dissipation scale; however, studies of turbulence at higher angular resolution than 30\arcsec should reveal this scale.

\subsection{Disk sizes}

Of the 12 Class 0 and I protostars in this region, dust emission observations of disks 
have resolved only six sources with either VLA or ALMA continuum observations \citep{Segura-Cox2016-Perseus_Disks_Pilot,Segura-Cox2018-Disk_Size,Diaz-Rodriguez2022-SVS13A_ALMA_highres}. 
The mean dust disk radii measured for these six resolved sources is $\approx$30 au (see Table~\ref{tab:Rdisk}). 
These values are lower than the typical values obtained for Class 0 and I
objects \citep[median values of 61 au and 81 au for Class 0 and I, respectively; see][for a recent review]{Tsukamoto2023-PP7_Disk_Formation}.

Theoretical models suggest disk properties are sensitive to the degree of coupling between ions and neutrals, with perfect coupling corresponding to the "ideal" MHD limit in which angular momentum is efficiently transported away from the circumstellar region by magnetic braking \citep{ZhaoTomida2020-Disk_Formation_Review}. Higher levels of CR ionization effectively enhance magnetic braking, thereby leading to smaller disks \citep{Dapp2012-Disk_Formation_AD_Ohm,Zhao2016nosmalldust-disk,Wurster2016}.
For example, \cite{Kuffmeier2020-Ionization_Disk_Size} carried out a suite 
of simulations that followed the collapse of spherical cores 
and showed that the disk size decreases with an increasing 
cosmic-ray ionization rate for identical 
initial conditions because of more efficient magnetic braking.
Although prestellar cores appear as more dynamic entities 
in larger scale simulations \citep{Kuffmeier2017-GMCtodisk,Lebreuilly2021,Offner2022-Core_Evolution}, 
the results from \cite{Kuffmeier2020-Ionization_Disk_Size} 
provide a well-controlled experiment with different 
mass-to-flux-ratios 
\begin{equation}
    \lambda_{mag} \equiv \frac{M_{\rm c} } {\pi R_{\rm c}^2 B_0} 2 \pi \sqrt{G}~,
\end{equation}
where $B_0$ is the magnetic field strength, and 
$M_{\rm c}$ and $R_{\rm c}$ are the mass and radius 
of the core. %

For the typical values of the ionization and cosmic rays found 
in this work, 
$\langle X(e) \rangle = 10^{-6.5}$ and 
$\langle \zeta(\ce{H2}) \rangle = 10^{-16.5}$ s$^{-1}$ 
= $3\times10^{-17}$ s$^{-1}$, 
the typical gas disk sizes found in this region are well matched 
by the cases in \cite{Kuffmeier2020-Ionization_Disk_Size} with 
$\lambda_{mag} = 5$ and an initial ratio of 
rotational to gravitational energy, $\beta$, of the 
spherical core between $\beta = 0.025$ and $\beta = 0.1$
(see also \cite{Wurster2018-ionization_disks}). 
Similarly, \cite{Lebreuilly2021} investigated the disk size 
distribution in nonideal-MHD models, assuming an ionization 
rate of $\langle \zeta(\ce{H2}) \rangle = 10^{-17}$ s$^{-1}$, 
of a collapsing clump of 1\,000 \msun. They found an 
average disk radius of $\approx$50 au, which is about a 
factor of two larger than what is found in our observations. 
This might indicate that the relatively higher 
cosmic-ray ionization rate in NGC 1333 
leads to smaller disk sizes, although a lower 
angular momentum or higher magnetization cannot be ruled out. 

We highlight that the cosmic-ray ionization rate is not 
a constant value throughout the entire region, and it may fluctuate, 
especially toward higher densities closer to the actively 
accreting protostars.
At these high densities, this rate could decrease due 
to the shielding of cosmic rays 
\citep{NakanoTademaru1972,UmebayashiNakano1981,Padovani+2009}, 
or it could increase due to local enhancement of cosmic-ray ionization 
by the individual protostars themselves \citep{Padovani+2016,GachesOffner2018}.

\subsection{Improvements on $X(e)$ and $\zeta(\ce{H2})$}
Although the method for determining the electron fraction and 
cosmic-ray ionization rate is widely used, there are clear 
paths for improvements.
The assumption of a single excitation temperature could impact 
the total column densities derived and the derived $X(e)$ and $\zeta(\ce{H2})$.
However, the \ce{H^{13}CO+} and \ce{DCO+} (1--0) lines 
have similar parameters ($E_{up}$ and $A_{ul}$), and 
therefore, this assumption is not expected to cause major issues.
Also, other transitions of these species, combined with an 
excitation analysis, could better constrain 
the optical depth and excitation temperatures of these species
and therefore derive more reliable molecular column densities.

Recently, \cite{Sabatini2023-High_mass_Zeta} has shown that in high-mass star-forming regions,
where \ce{DCO+} and \ce{H^{13}CO+} might trace different volumes or be affected by 
high optical depths, \ce{H2D+} is a better estimator for the cosmic-ray ionization rate.
Unfortunately, this approach is severely limited in low-mass star-forming regions 
since the \ce{H2D+} emission is compact and only traces the 
highest density regions in dense cores 
\citep[e.g.,][]{Caselli2003-H2Dp_L1544,vanderTak2005-Line_Profiles_L1544,Harju2006-H2Dp_OrionB,Vastel2006-H2Dp_L1544,Caselli2008-H2Dp_Survey,Friesen2010-OphB2_Deuteration,Friesen2014-OphA_ALMA_H2Dp,Koumpia_2020}.

A complementary approach might involve analyzing multiple molecular 
species and/or transitions and comparing the line emission
with a suite of chemical models to directly constrain the 
typical physical parameters of the region 
(e.g., density, radiation field, $\zeta(\ce{H2})$, "chemical" age). 
This could be carried out with single-zone models (0D) or 
more complex ones using some physical model for the region. 
Regardless of the approach, a new and exciting possibility of 
deriving robust and high-quality resolved maps of 
$X(e)$ and $\zeta(\ce{H2})$ is upon us.

\section{Conclusions}

We have presented the first results of the ProPStar survey. 
We studied the southeast section of the NGC 1333 region using 
combined observations of the 30-m and NOEMA telescopes to recover all scales. 
The main results are listed below:
\begin{enumerate}
  \item We mapped the \ce{DCO+} and \ce{H^{13}CO+} (1--0)
      emission at $\approx 6\arcsec$ resolution 
      ($\approx$1\,800 au).
  \item We used the analytic relation from \cite{Caselli_1998-Ion_Fraction} 
      relating the measured 
      \ce{DCO+} and \ce{H^{13}CO+} column density ratio 
      in addition to archival \ce{C^{18}O} and \ce{H2} in order
      to derive the first large-area map for the electron 
      fraction, $X(e)$, and cosmic-ray ionization rate, $\zeta(\ce{H2})$,
      in a low-mass star-forming region.
  \item The electron fraction and cosmic-ray ionization rate distributions 
  {%
  are peaked at about} $10^{-6.5}$ and $10^{-16.5}$ s$^{-1}$, respectively,
  {%
  showing} local enhancements around the embedded YSOs. These values are higher 
  than expected from the "standard" cosmic-ray ionization rate in 
  molecular clouds.
  \item The northwest region in the map shows highly elevated values for both the 
       electron fraction and cosmic-ray ionization rate, which might be a
       sign of the interaction between the molecular cloud and a local bubble, 
       or it could be due to cosmic rays being locally generated by YSOs.
  \item We used the typical value of $X(e)$ in the region to estimate the 
  critical length for wave propagation of MHD turbulence of 0.054 pc ($\approx$37\arcsec). 
  If the turbulence in the molecular clouds is of MHD nature, then this dissipation 
  scale could be
  measured with observations of enough sensitivity and resolution.
  \item The increased values of $X(e)$ and $\zeta(\ce{H2})$ are sufficiently high such
  that when compared to numerical simulations, the newly formed disks should be 
  consistent with ideal-MHD simulations (e.g., small). This also suggests that 
  disk formation in clustered regions could strongly depend on the feedback 
  from the first stars that are formed.
\end{enumerate}

In summary, these observations show the great potential offered by 
combined NOEMA and IRAM 30-m observations to determine the electron fraction and 
cosmic-ray ionization rate in the NGC 1333 region. 
Both of these quantities are important to understanding the physical conditions 
and processes involved in star and disk formation.
More maps of the cosmic-ray ionization rate and electron fraction, combined with complementary 
disk size measurements, should provide a much better understanding of 
the role of stellar feedback (e.g., locally generated cosmic rays) 
and the importance of nonideal MHD in the whole star- and disk-formation processes.

\begin{acknowledgements}
    The authors kindly thank the anonymous referee for the comments that helped improve the manuscript.
    Part of this work was supported by the Max-Planck Society. 
    DMS is supported by an NSF Astronomy and Astrophysics Postdoctoral Fellowship under award AST-2102405.
    We gratefully acknowledge the helpful discussions with Elena Redaelli, 
    Cecilia Ceccarelli, Giovanni Sabatini, and Sylvie Cabrit,
    which improved the article.
    We thank Elena Redaelli for providing the electronic version 
    of the electron fraction and density data points from 
    modeling L1544. 
    We thank Yasuo Doi for providing the electronic version 
    of the polarization vectors.
    This research has made use of NASA’s Astrophysics Data System.
\end{acknowledgements}

\bibliographystyle{aa}
\bibliography{NGC1333_Electron_Fraction_map}

\begin{appendix}

\section{Optical depth of \ce{C^{18}O}\label{appendix:tau}}
We estimated the \ce{C^{18}O} (3--2) line optical depth as
\begin{equation}
\tau(\ce{C^{18}O}) = 
-\ln\left(1.0 - \frac{T_p (\ce{C^{18}O})} 
                   {J_{\nu}(T_{ex}) - J_\nu(T_{cmb})} \right)~,
\end{equation}
where $T_p (\ce{C^{18}O})$ is the peak brightness of the line, and 
$J_\nu(T)$ is the brightness temperature of a black body 
with temperature $T$ at frequency $\nu$. In the case of \ce{C^{18}O} (3--2),
we assumed $T_{ex} = 12$~K.

Figure~\ref{fig:tau_c18o} shows the map of the optical depth.
The figure clearly shows that the emission is mostly optically thin, which is also consistent with the  
findings from \cite{Curtis2010-JCMT_Perseus_CO}.

\begin{figure}[ht]
    \centering
    \includegraphics[width=\columnwidth]{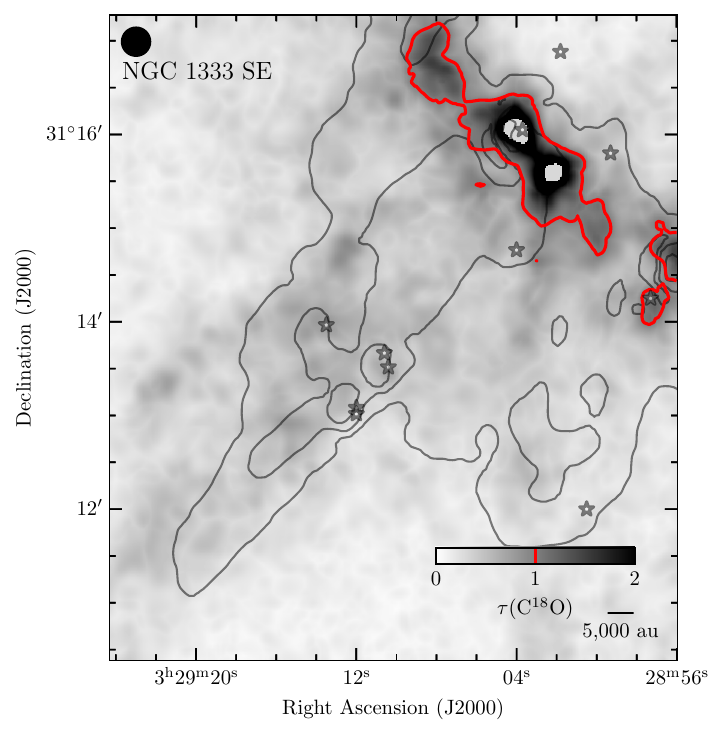}
    \caption{Optical depth of \ce{C^{18}O} (3--2). 
    The red line shows the $\tau(\ce{C^{18}O})=1$ contour.
    The contours correspond to the {\it Herschel}-based 
    \ce{H2} column map; the contours are the same as in Fig.~\ref{fig:maps_TdV}.
    The beam size and scale bar are shown in the top-let and bottom-right corners, respectively.}
    \label{fig:tau_c18o}
\end{figure}

\section{The effect of excitation temperature}
One of the most important simplifications in this work was the use of a single excitation temperature for the column 
density calculation of \ce{H^{13}CO+} and \ce{DCO+}, which we used to determine the $R_{\rm D}$ and $R_{\rm H}$ quantities.
We explored the effect of both species having the same excitation temperature 
but that temperature being different from 10 K (5 K $< T_{\rm ex} < $ 12 K), and we found that 
the variation in $R_{\rm D}$ is less than $5\%$, while for $R_{\rm H}$, the variation is less than $10\%$.
Therefore, an improved excitation analysis of both species would provide a substantial check
to the calculation used in this work.

{\section{$\Delta \log_{10} \zeta$\label{Appendix:delta_zeta}}
We calculated the difference between the derived CRIR and the expected values from the $\mathcal{L}$ model of CRIR as a function of column density from 
\cite{Padovani2018-Cosmic_Ray_Models}, 
 $\Delta \log_{10}  \zeta = \log_{10}  \zeta_{obs} - \log_{10}  \zeta_{model}$. 
Figure~\ref{fig:delta_zeta} shows the map of $\Delta \log_{10}  \zeta$, which is similar to Fig.~\ref{fig:zeta}, with lower values than expected in the southeast section,
local increases toward YSOs, and a general excess in the rest of the map.
}

\begin{figure}[ht]
    \centering
    \includegraphics[width=\columnwidth]{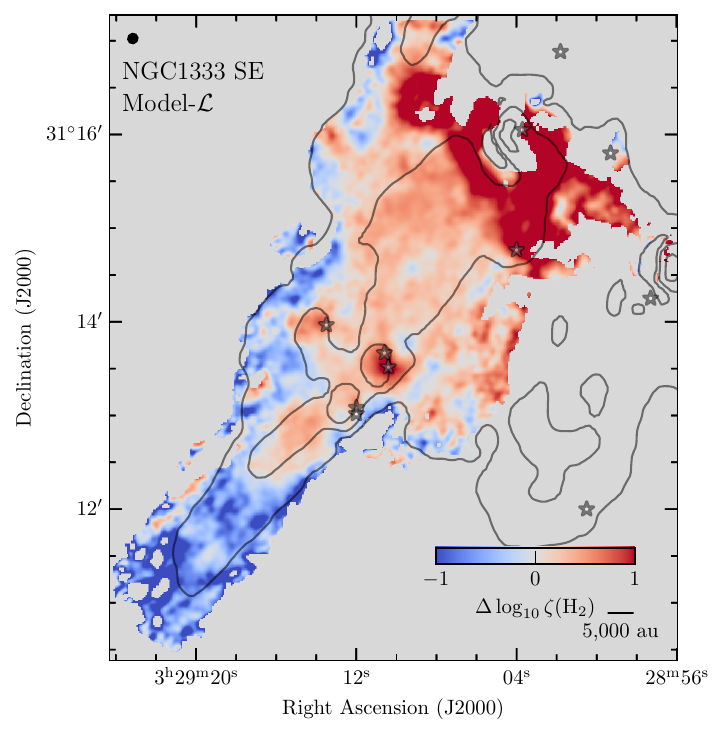}
    \caption{{Difference between observed CRIR and the model-$\mathcal{L}$ to better illustrate the local effects,
    $\Delta \log_{10}  \zeta = \log_{10}  \zeta_{obs} - \log_{10}  \zeta_{model}$. 
    The red and blue colors show the excess and deficiency of CRIR with respect to the expected value 
    following \cite{Padovani2018-Cosmic_Ray_Models}. 
    The contours correspond to the {\it Herschel}-based 
    \ce{H2} column map; the contours are the same as in Fig.~\ref{fig:maps_TdV}.
    The beam size and scale bar are shown in the top-let and bottom-right corners, respectively.}}
    \label{fig:delta_zeta}
\end{figure}

\section{Disk sizes and outflow directions of covered YSOs}
We list the source name, disk radii, and outflow directions found in the literature 
in Table~\ref{tab:Rdisk}. Except for SVS13 VLA4A and SVS13 VLA4B, the disk sizes are the result of modeling VLA 8.1~mm continuum observations, which have a FWHM beam size of $\sim$16 au \citep{Segura-Cox2016-Perseus_Disks_Pilot, Segura-Cox2018-Disk_Size}.  For sources without a reported disk radii, the disks were detected but unresolved in \citet{Segura-Cox2018-Disk_Size}, and the disk radii at VLA wavelengths could only be estimated as being less than or equal to 8 au.  For SVS13 VLA4A and SVS13 VLA4B, which form a close binary that shares dusty circumbinary material, the dust disk radii of the individual circumstellar disks are reported from Gaussian fits to ALMA 0.9~mm continuum observations, with a FWHM beam size of $\sim$26 au \citep{Diaz-Rodriguez2022-SVS13A_ALMA_highres}.

\begin{table*}[htb]
\caption{Properties of Class 0 and I protostars within the field of view.\label{tab:Rdisk}}
\begin{tabular}{lccccc}
\hline \hline
Source & Class & $R_{disk}$\tablefootmark{a} & PA$_{outflow}$\tablefootmark{b} & Other Name & Ref. \\
 & & (au) & (deg) & \\
 \hline
NGC 1333 IRAS4A1\tablefootmark{c} & 0 & 49.7  & 19.2 & Per-emb-12  & (1, 5) \\ 
NGC 1333 IRAS4A2\tablefootmark{c} & 0 & --  & 20 & Per-emb-12  & (1, 7) \\ 
NGC 1333 IRAS4B\tablefootmark{c} & 0 & --  & 176 & Per-emb-13 & (1, 2, 5)\\ 
NGC 1333 IRAS4B2\tablefootmark{c} & 0 & -- & 76 & Per-emb-13 & (1, 8)\\  
NGC 1333 IRAS2B & I & -- & 24 & Per-emb-36 & (1, 9)  \\ 
NGC 1333 IRAS4C & 0 & 37.2 & 96 & Per-emb-14 & (2, 6)\\ 
SVS13 VLA4A\tablefootmark{d} & 0/I & 12  & 8 & SVS13A, Per-emb-44  & (3, 4)\\ 
SVS13 VLA4B\tablefootmark{d} & 0/I & 9  & -50 & SVS13A, Per-emb-44  & (3, 4)\\ 
SVS13 VLA17 & 0 & 31.7 & 170 & SVS13B & (2, 5)\\ 
SVS13C & 0 & 42.9 & 8 & -- & (1, 4)\\ 
Per-emb-15 & 0 & -- & -35 & SK14 & (4) \\ 
Per-emb-3 & 0 & -- & 97 & -- &  (1, 9)\\ 
\hline
\end{tabular}
\tablebib{
(1) \cite{Segura-Cox2018-Disk_Size},
(2) \cite{Segura-Cox2016-Perseus_Disks_Pilot},
(3) \cite{Diaz-Rodriguez2022-SVS13A_ALMA_highres},
(4) \cite{Plunkett2013-CARMA_NGC1333},
(5) \cite{Lee2016-MASSES_Outflow_Pilot},
(6) \cite{Zhang2018-PEACHES_IRAS4C_Outflow},
(7) \cite{Chuang2021_IRAS4A2},
(8) \cite{Hull2014_TADPOL},
(9) \cite{Stephens2017-MASSES_Outflows_Angles}
}
\tablefoot{
\tablefoottext{a}{The disk radius measurements are dust based, and they were recomputed to the same 300 pc distance, if needed.}
\tablefoottext{b}{Outflow PA was measured using the red lobe from due east from north.}
\tablefoottext{c}{These objects form a binary system but do not share circumbinary material.}
\tablefoottext{d}{These objects belong to a close binary system 
surrounded by circumbinary material or disk.} %
}
\end{table*}

\end{appendix}
\end{document}